# Testing Local Lorentz and Position Invariance and Variation of Fundamental Constants by searching the Derivative of the Comparison Frequency Between a Cryogenic Sapphire Oscillator and Hydrogen Maser

Michael Edmund Tobar,[1] Peter Wolf,[2] Sébastien Bize,[2] Giorgio Santarelli,[2] Victor Flambaum[3]

[1]School of Physics, University of Western Australia, Crawley, Western Australia
[2]LNE-SYRTE, Observatoire de Paris, CNRS, UPMC, 75014 Paris, France
[3]School of Physics, The University of New South Wales, Sydney NSW, Australia

The cryogenic sapphire oscillator (CSO) at the Paris Observatory has been continuously compared to various Hydrogen Masers since 2001. The early data sets were used to test Local Lorentz Invariance in the Robertson-Mansouri-Sexl (RMS) framework by searching for sidereal modulations with respect to the Cosmic Microwave Background, and represent the best Kennedy-Thorndike experiment to date. In this work we present continuous operation over a period of greater than six years from September 2002 to December 2008 and present a more precise way to analyse the data by searching the time derivative of the comparison frequency. Due to the long-term operation we are able to search both sidereal and annual modulations. The results gives $P_{KT} = \beta_{RMS} - \alpha_{RMS} - 1 = -1.7(4.0) \times 10^{-8}$ for the sidereal and $-23(10) \times 10^{-8}$ for the annual term, with a weighted mean of $-4.8(3.7) \times 10^{-8}$, a factor of 8 better than previous. Also, we analyse the data with respect to a change in gravitational potential for both diurnal and annual variations. The result gives $\beta_{H\text{-}Maser} - \beta_{CSO} = -2.7(1.4) \times 10^{-4}$ for the annual and $-6.9(4.0) \times 10^{-4}$ for the diurnal terms, with a weighted mean of $-3.2(1.3) \times 10^{-4}$. This result is two orders of magnitude better than other tests that use electromagnetic resonators. With respect to fundamental constants a limit can be provided on the variation with ambient gravitational potential and boost of a combination of the fine structure constant ($\alpha$), the normalized quark mass ($m_q$), and the electron to proton mass ratio ($m_e/m_p$), setting the first limit on boost dependence of order $10^{-10}$.

## I INTRODUCTION

Local Lorentz Invariance (LLI) and Local Position Invariance (LPI) are underlying principles of relativity and Einstein's Equivalence Principe (EEP). LLI postulates that the outcome of a local experiment is independent of the velocity and orientation of the apparatus, while LPI implies that the gravitational redshift of radiation is universal and thus independent of the type of emitting source and that fundamental constants remain constant with respect to time and space. Tests of EEP are motivated by its central importance to modern physics, as well as the development of a number of unification theories, which suggest violations of EEP at some level. To identify a violation it is necessary to have an alternative test theory to interpret the experiment[1], which have been developed for both LLI experiments[2-8] and LPI experiments[1,9,10].

## A. Local Lorentz Invariance

For tests of LLI the kinematical frameworks (RMS)[2,3] postulate a simple parameterization of the Lorentz transformations with experiments setting limits on the deviation of those parameters from their values in special relativity. Due to their simplicity they have been widely used to interpret many experiments[11-16]. In general these tests are performed with respect to the Cosmic Microwave Background (CMB), which is considered as our best candidate for an absolute reference frame.

More recently, a general Lorentz violating extension of the standard model of particle physics (SME) has been developed whose Lagrangian includes all parameterized Lorentz violating terms that can be formed from known fields[6-8,17]. This has inspired a new wave of experiments designed to explore uncharted regions of the SME Lorentz violating parameter space. Because of the vast amount of parameters there has been an increase in activity in experimental tests of LLI, in particular light speed isotropy tests (or Michelson-Morley experiments) with at least 7 experiments reported in the last 6 years[15,18-26] as well as new Ives-Stillwell experiments[27-30]. This is largely due to advances in technology, allowing more precise measurements, and the emergence of the Standard Model Extension (SME) as a general framework for the analysis of experiments, providing new interpretations of LLI tests. It has been recently shown that the SME fully contains the RMS framework and is thus more general[31]. None of these experiments have yet reported a violation of LLI, though the constraints on a putative violation have become more stringent by up to three orders of magnitude in the same time frame.

*Testing Boost Dependence*

LLI experiments with respect to boost, such as Kennedy-Thorndike and Standard Model Extension (SME) tests must be performed with respect to a specific frame of reference. Since the discovery of the Cosmic Microwave Background (CMB) frame of reference[32], historically the former types of experiments have been undertaken with respect to this frame. Alternatively, SME tests have been done with respect to a sun-centred frame, which remains inertial with respect to the CMB. Our earlier data has been used to test for boost and isotropy components in the photon sector of the SME as well as attaining the best limit for the Kennedy Thorndike experiment with respect to the CMB[14,15,19]. For the SME, rotating experiments have superseded this experiment[16,20,23,25], but remain non-sensitive Kennedy-Thorndike experiments. An earlier version of this experiment still remains the most sensitive Kennedy-Thorndike test. However, we now have more than six years of continuous data and are able to put better limits. Also, a putative time or red shift dependence of fundamental constants implies that on a cosmic scale there may also manifest a boost dependence with respect to the preferred frame[33,34]. In this work we put limits on the boost dependence of fundamental constants with respect to the CMB and redo the Kennedy-Thorndike experiment with a factor eight better sensitivity.

To calculate the boost with respect to the CMB we transform all quantities to a geocentric non-rotating reference frame (with respect to distant stars) centered at the centre of mass of the Earth (Earth frame) with its z-axis perpendicular to the equator, pointing north, the x-axis pointing towards 11.2h right ascension (aligned with the equatorial projection of $u$ defined below in (1)). A pictorial representation of the frame is shown in fig. 1. Classical (Galilean) transformations for the velocities are sufficient as relativistic terms are of order $O(c^{-2})$. We consider three velocities: the velocity of the sun with respect to the cosmic microwave background $u$ (declination -6.4°, right ascension 11.2 h), the orbital velocity of the Earth $v_0$ and the velocity of the lab due to the rotation of the Earth $v_R$. The sum of those three will provide the velocity of the laboratory.

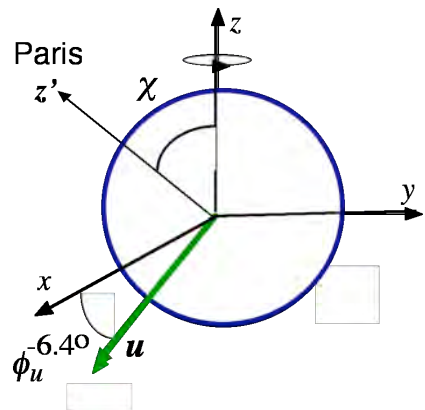

Fig. 1. The Earth-centered frame we use has the spin axis along the z-axis, and the velocity of the Sun with respect to the CMB is defined to have no component in the y direction. Thus, the Earth is spinning at the sidereal rate within this frame. The angle χ of Paris with respect to the z-axis is shown pictorially.

In the Earth frame the CMB velocity is:

$$\mathbf{u} = u \begin{bmatrix} \cos\phi_\mu \\ 0 \\ \sin\phi_\mu \end{bmatrix} \quad (1)$$

where, $u$ is approximately 377 km/s and $\phi_u$ is approximately -6.4°. The orbital velocity with respect to the Earth frame has been shown to be[35]:

$$\mathbf{v}_O = v_O \begin{bmatrix} -\sin\lambda_o \cos\eta + \cos\lambda_o \sin\eta \cos\varepsilon \\ \sin\lambda_o \sin\eta + \cos\lambda_o \cos\eta \cos\varepsilon \\ \cos\lambda_o \sin\varepsilon \end{bmatrix} \qquad (2)$$

Here, $v_o$ is the orbital speed of the Earth, which is approximately 29.78 km/s, $\varepsilon$ the angle between the equatorial and orbital planes is approximately 23.27°, $\eta$ the right ascension of $\mathbf{u}$ is approximately 167.9° and $\lambda_o = \Omega_\oplus(t - t_o)$ where $t-t_o$ is the time since the J2000 equinox (20 March at 07 h 35 min UTC for 2000) and $\Omega_\oplus$ is the annual frequency (0.017203 rads/day).

The velocity at Paris due to the rotation of the Earth is given by:

$$\mathbf{v}_R = \omega_\oplus R \cos\chi \begin{bmatrix} -\sin\lambda \\ \cos\lambda \\ 0 \end{bmatrix} \qquad (3)$$

Here, $\omega_\oplus$ is the angular frequency of the Earth rotation with respect to the stars (sidereal frequency $7.29 \times 10^{-5}$ rads/s), $R$ is the equatorial radius of the Earth (6,370 km), $\chi$ is the latitude (48.7° for Paris) and $\lambda = \omega_\oplus t + \Phi$ is the longitude in the Earth frame. The phase $\Phi$ is set by the initial longitude for Paris at time $t = 0$ defined by the J2000 equinox. Thus the total boost ($\mathbf{b}$) with respect to the CMB is given by:

$$\mathbf{b} = \frac{\mathbf{v}_T}{c} = \frac{\mathbf{u} + \mathbf{v}_O + \mathbf{v}_R}{c} \qquad (4)$$

Substituting all values into (4) one obtains the boost vector:

$$\mathbf{b}(t) = \begin{bmatrix} 1.25 \times 10^{-3} + 1.91 \times 10^{-5} \cos[\Omega_\oplus(t-t_o)] + 9.71 \times 10^{-5} \sin[\Omega_\oplus(t-t_o)] - 1.02 \times 10^{-6} \sin[\omega_\oplus(t-t_o) + \Phi] \\ -8.92 \times 10^{-5} \cos[\Omega_\oplus(t-t_o)] + 2.08 \times 10^{-5} \sin[\Omega_\oplus(t-t_o)] + 1.02 \times 10^{-6} \cos[\omega_\oplus(t-t_o) + \Phi] \\ -1.40 \times 10^{-4} + 3.92 \times 10^{-5} \cos[\Omega_\oplus(t-t_o)] \end{bmatrix} \qquad (5)$$

The boost vector given by (5) shows both annual and sidereal variation and in the following work we search for Lorentz violations and fundamental constant variation with respect to the boost. To search for leading order variations the boost vector (5) is used and for second order terms the magnitude squared, only keeping the dominant quadrature terms from (5) we calculate:

$$|\mathbf{b}| \approx 1.3 \times 10^{-3} + 1.5 \times 10^{-5} \cos[\Omega_\oplus(t-t_o)] + 9.7 \times 10^{-5} \sin[\Omega_\oplus(t-t_o)] \\ -1.0 \times 10^{-6} \sin[\omega_\oplus(t-t_o) + \Phi] - 7.6 \times 10^{-8} \cos[(\omega_\oplus - \Omega_\oplus)(t-t_o) + \Phi] \qquad (6)$$

$$|\mathbf{b}|^2 \approx 1.6 \times 10^{-6} + 3.7 \times 10^{-8} \cos[\Omega_\oplus(t-t_o)] + 2.4 \times 10^{-7} \sin[\Omega_\oplus(t-t_o)] \\ -2.6 \times 10^{-9} \sin[\omega_\oplus(t-t_o) + \Phi] - 1.9 \times 10^{-10} \cos[(\omega_\oplus - \Omega_\oplus)(t-t_o) + \Phi] \qquad (7)$$

The dominant terms in (5) to (7) above are at the annual, sidereal and diurnal ($\omega_\oplus - \Omega_\oplus$) frequencies. For the Kennedy-Thorndike experiment we search variations in the RMS framework ($P_{KT} = \beta_{RMS} - \alpha_{RMS} - 1$, see[15] for the derivation) from the comparison between the H-maser and CSO frequency of the form;

$$\frac{\Delta\nu_{CSO-Hmaser}(t)}{\nu} = P_{KT}|b(t)|^2 \qquad (8)$$

Thus, our approach in undertaking the Kennedy-Thorndike experiment is to search from the data the quadrature amplitudes at the frequencies given in (7) to put a limit on $P_{KT}$.

### B. Local Position Invariance

For LPI measurements a simple parameterization with respect to change in gravitational potential ($\Delta U$) has been made based on different clocks (denoted by subscript $i$), which is given by[12,36-41],

$$\frac{\Delta\nu_i(t)}{\nu} = (1+\beta_i)\frac{\Delta U(t)}{c^2} \qquad (9)$$

Here $\Delta\nu_i/\nu$ is the apparent fractional frequency difference of clock $i$ with respect to potential, and the parameter $\beta_i$ will depend on the type of clock. For example, an ion or atomic clock will depend on the species and type of transition, while for a classical oscillator dependence on dimensional and refractive index changes may be considered. For a null redshift experiment a comparison of two different types of clocks is necessary, and from (9) we may write:

$$\frac{\Delta\nu_{a-b}(t)}{\nu} = \frac{\nu_a(t)-\nu_b(t)}{\nu} = (\beta_a - \beta_b)\frac{U(t)}{c^2} \qquad (10)$$

Here $\Delta\nu_{a-b}/\nu$ is the frequency difference between two different types of clocks that are co-located.

On earth we consider the changing potential due to the solar gravitational potential, so that

$$\frac{U(t)}{c^2} = -\frac{Gm_s}{ac^2}e\cos(\Omega_\oplus t_\oplus) \qquad (11)$$

Here G is the gravitational constant, $m_s$ the mass of the sun, $a$ the semi major axis, $e$ the eccentricity of the orbit, $\Omega_\oplus$ the angular frequency of a sidereal year and $t_\oplus$ the time elapsed with respect to a recorded Aphelion. The expected change in gravitational potential given by (11) is determined by fitting the earth's orbit around the sun from the table of Perihelion and Aphelion data (2000-20)[42], with phase and normalised amplitude (by factor $-\frac{Gm_s}{ac^2}e \approx -1.65 \times 10^{-10}$) shown in fig. 2.

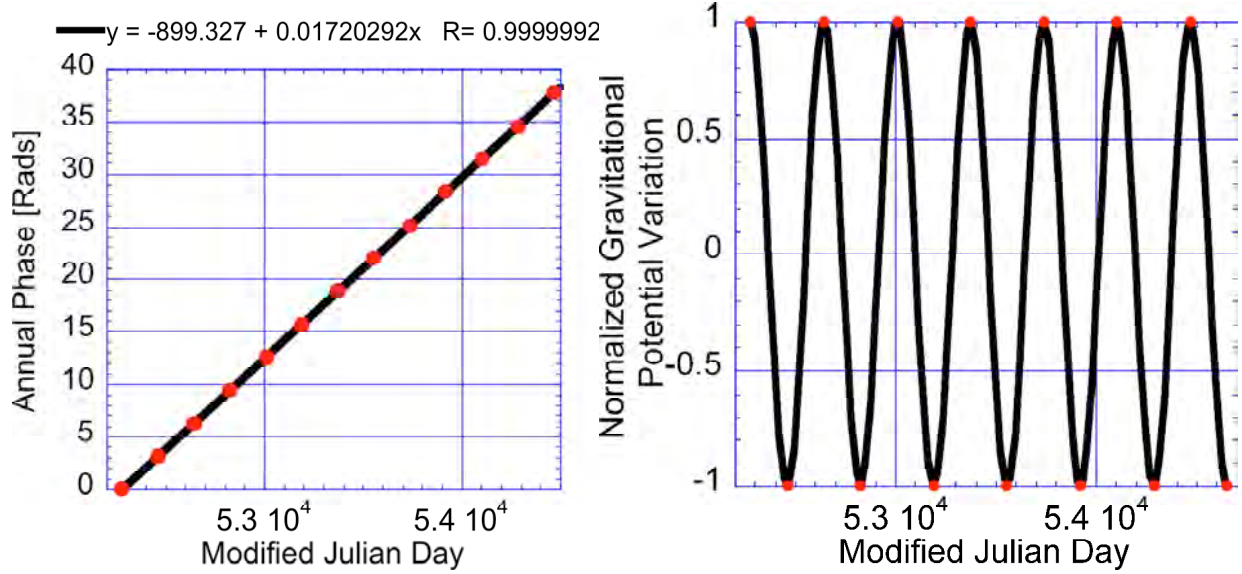

Fig. 2. Right, normalized change in gravitational potential of the Sun at the Earth, due to the Earth's orbital eccentricity. Left, phase of the gravitational potential versus Modified Julian Day (MJD). The slope gives the frequency of the signal as 0.017203 rads/day (the annual frequency).

Diurnal variations will also occur as the Earth rotates from day to night. The phase is set when Paris is directly pointing towards the Sun. This phase can be calculated with respect to midday Greenwich, which is also midday for a Modified Julian Day. The longitude of Paris is 2.33°, thus the phase is advanced by 0.0406662 radians with respect to midday of a Modified Julian Day. Noise statistics of the CSO is better over a day when compared to a year so tests of both LPI and LLI can be compared over these two time scales. The magnitude of the signal (variation of gravitational potential) is to first order sinusoidal with an amplitude approximately given by; $\frac{Gm_s r_{paris}}{c^2 a^2} \approx 2.54 \times 10^{-13}$, $r_{paris} = r_{Earth} Cos[\phi_{paris}] Cos[\psi_{Earth}]$, where $r_{Earth}$ is $6.37 \times 10^6$ meters and $\phi_{paris}$ is the latitude (48.7°) and $\psi_{Earth}$ is the Earth's tilt of 23.5°.

### C. Variation With Respect to Fundamental Constants

It has also been suggested that variation in fundamental constants can occur with respect to space and time, with respect to the evolving universe[43]. Effects proportional to gravity may occur through couplings of the standard model fields to an evolving scalar field ($\phi$) such as the dilaton[44,45], which is proportional to the gravitational potential given by (12).

$$\phi - \phi_0 = \kappa \left( GM/rc^2 \right) \qquad (12)$$

Thus, the more recent approach to testing LPI relates the dependence of the type of frequency standard to fundamental constants[39-41,46,47]. This approach includes calculating how the frequency-determining element depends on dimensionless constants such as the fine structure constant, $\alpha$, the quark mass ($m_q$) with respect to the quantum chromodynamics (QCD) scale of the strong force ($\Lambda_{QCD}$), $\mu_q = m_q/\Lambda_{QCD}$ and the electron to proton mass ratio $\mu_e = m_e/m_p \propto m_e/\Lambda_{QCD}$. In general the frequency ratio between two different clocks (or frequency standards) with respect to dependence of fundamental constants is of the form;

$$x = \frac{\nu_1}{\nu_2} = Const \times \alpha^{n_1} \mu_e^{n_2} \mu_q^{n_3} \tag{13}$$

The values of $n_1$, $n_2$ and $n_3$ depend on the frequency standards used in the comparison. Thus, any variation of the fundamental constants with respect to gravitational potential will induce a frequency shift of the form;

$$\frac{\delta x}{x} = (n_1 \kappa_\alpha + n_2 \kappa_e + n_3 \kappa_q) \times \delta\left(\frac{GM}{rc^2}\right);$$

$$\kappa_\alpha = \frac{\delta\alpha/\alpha}{\delta\left(\frac{GM}{rc^2}\right)}; \quad \kappa_e = \frac{\delta\mu_e/\mu_e}{\delta\left(\frac{GM}{rc^2}\right)}; \quad \kappa_q = \frac{\delta\mu_q/\mu_q}{\delta\left(\frac{GM}{rc^2}\right)} \tag{14}$$

In the laboratory, one can look for variations with respect to the changing gravitational potential in the same way as the LPI experiment discussed previously. For comparisons between the CSO and H-maser the values of $n_1$, $n_2$ and $n_3$ are 3, 1 and -0.1 respectively[36,48], unless the frequency is close to a phonon or paramagnetic resonance (in this case sensitivity would be enhanced[49]), which is not the case in our experiment.

It has also been shown that fundamental constant may vary through the association of Lorentz and CPT symmetry violations[50]. Thus, one may also expect fundamental constants to vary with respect to boost as well as gravitational potential. Furthermore, variations in fundamental constants have been postulated to occur through a scalar field that is dependent on redshift[34]. One interpretation of the redshift dependence is a variation with respect to look back time (time dependence). On the other hand, one could postulate a variation due to the direct boost. If one is to search for boost dependence of fundamental constants it makes more sense to search for effects relative to the preferred frame candidate the Cosmic Microwave Background to complement the cosmological redshift searches. Stringent limits on fundamental constants have been set on cosmological scales with respect to the cosmological redshift[51-55]. However, frequency comparisons experiments offer a more sensitive measurement if the two frequencies have a different dependence on fundamental constants. From (13) one can then postulate the dependence on boost to be of the form;

$$\frac{\delta x}{x} = \mathbf{B} \cdot \mathbf{b}; \quad \mathbf{B} = n_1 \mathbf{B}_\alpha + n_2 \mathbf{B}_e + n_3 \mathbf{B}_q$$

$$\mathbf{B}_{\alpha_i} = \frac{1}{\alpha} \frac{\delta \alpha}{\delta b_i}; \quad \mathbf{B}_{e_i} = \frac{1}{\mu_e} \frac{\delta \mu_e}{\delta b_i}; \quad \mathbf{B}_{q_i} = \frac{1}{\mu_q} \frac{\delta \mu_q}{\delta b_i}$$

(15)

Here, the violation due to the boost is described by the vectors $\mathbf{B}_\alpha$, $\mathbf{B}_e$ and $\mathbf{B}_q$ with components $i = x, y$ or $z$.

Thus, in the laboratory one may look for variations with respect to our changing boost over the year and sidereal day, as has been undertaken in the past for Kennedy Thorndike and SME type LLI experiments.

## II FREQUENCY COMPARISON EXPERIMENT

In this work we present near continuous monitoring of the CSO with respect to a H-maser from September 2002 to December 2008. Both oscillators are operated in temperature controlled rooms, with the temperature sensitive electronics mounted on an actively temperature stabilized panel. The CSO resonant frequency at 11.932 GHz is compared to the 100 MHz output of the hydrogen maser as shown as described in[55-57]. The maser signal is multiplied up to 12 GHz of which the CSO signal is subtracted. The remaining 67 MHz signal is mixed to a synthesizer signal at the same frequency and the low frequency beat at of 63-67 Hz is counted, giving access to the frequency difference between the maser and the CSO with respect to 12 GHz.

We analyse the data with respect sidereal, diurnal and annual modulations using weighted (WLS) or ordinary (OLS) least squares depending on the noise type (early data before September 2002 was discarded due to lack of environmental temperature control). To search for annual variations the data sets are averaged over 95,000 sec intervals (1.1 days) minimum data separation. To search for sidereal or diurnal variations the data sets are averaged over 2,500 second intervals (42 minutes) minimum data separation. We show that there is an advantage to analysing the derivative of the beat frequency over the beat frequency directly, because it naturally filters out non-stationary effects such as systematic jumps between cryogenic refills and relocking of the CSO leading to a more sensitive measurement. When fitting annual modulations to the data, the derivative gives a factor of four better sensitivity than the direct beat. Conversely, over sidereal and diurnal time scales the beat and the derivative give similar results. However, this is only attained after a rigorous conditioning of the beat frequency data, where one has to try and eliminate the non-stationary jumps in frequency one at a time by breaking up the data sets into subset, which does not include any noticeable jumps.

Thus, the data conditioning over such a large amount of data is a time consuming process. In contrast, the derivative is easily conditioned as frequency jumps simply appear as outliers reducing their effect on the fitted amplitudes. Thus, in this work we only set limits on putative signals with respect to the derivative of the beat frequency.

## III TESTING LORENTZ INVARIANCE WITH RESPECT TO BOOST

### A. Search for Sidereal and Diurnal Variations

The experimental data consists of 212 separate sets of measured values of the beat frequency (every 100 s) of varying lengths ranging from no less than 1 day to 24.5 days (average 7.6 days) spanning 2296 days (6.3 years) with a duty cycle of 71%. Periods of down time have occurred due to oscillator warm up for maintenance. Also, data sets are chosen to be free of excursions due to settling after cool downs and liquid helium top ups, which cause an exponential relaxation of the beat frequencies that take about a day to settle. Such data is vetoed as the frequency variation is dominated by a known systematic effect and adds to the reduction of the duty cycle of the measurement. Each data set is averaged over at least 2,500 second periods (size of the data set reduced by a factor of 25 from the original 100s). The residuals of the experimental data are shown in fig. 3 (after removing an offset and drift from all 212 sets).

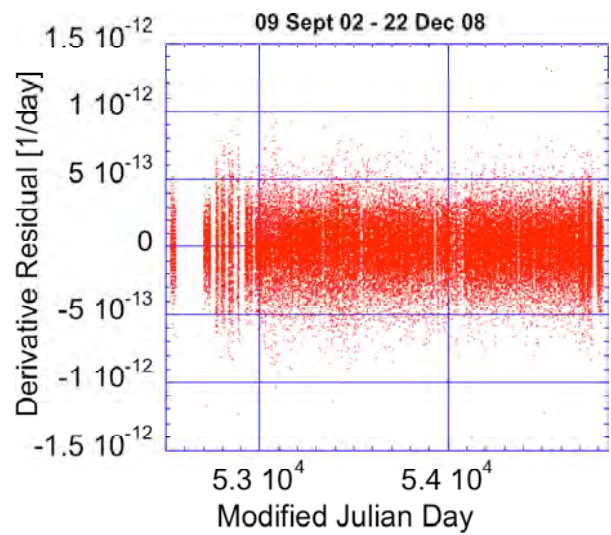

Fig.3. Fractional residuals of the derivative of the beat frequency (with respect to 12 GHz) of the 212 data sets.

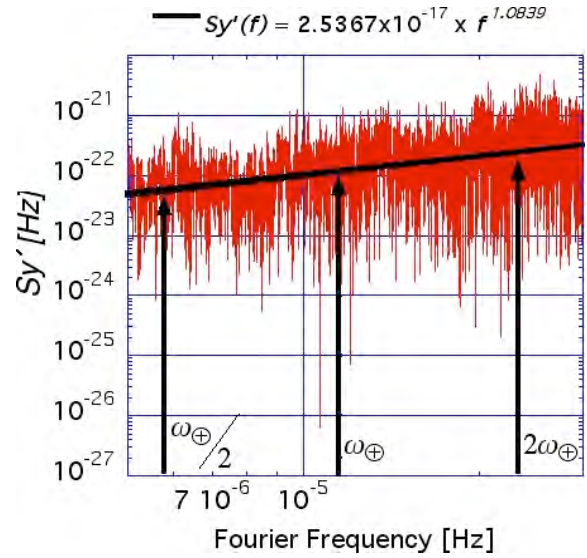

Fig. 4. Power spectral density of the residuals of the derivative of the fractional frequency fluctuations (Fig 4) around the frequencies of interest. The bold line is the power law fit.

The power spectral density of the residuals shown in Fig. 3 was observed to have a power law of $f^1$ around the frequencies of interest (consistent with prior analysis on the beat frequency[14]) as shown in Fig. 4. Thus, WLS is used to whiten the data[15], which optimises the search for significant frequency components with periods of the order of a day. With respect to the beat frequency given by eqn. (8) we search for the dominant components given by, $C_{\omega_\oplus - \Omega_\oplus} Cos[(\omega_\oplus - \Omega_\oplus)(t - t_o)]$ (in phase diurnal component), and $S_{\omega_\oplus} Sin[\omega_\oplus(t - t_o)]$ (out of phase sidereal component) with respect to the CMB. However, since we are searching for signals amongst the residuals of the derivative of the beat frequency, the putative signal will be differentiated and of the form, $\omega_\oplus S_{\omega_\oplus} Cos[\omega_\oplus(t - t_o)]$ and $-(\omega_\oplus - \Omega_\oplus)C_{\omega_\oplus - \Omega_\oplus} Sin[(\omega_\oplus - \Omega_\oplus)(t - t_o)]$ respectively. In actual fact we simultaneously fit to the residuals 40 parameters given by, $\sum_i -\omega_i C_{\omega i} Sin[\omega_i(t - t_o)] + \omega_i S_{\omega i} Cos[\omega_i(t - t_o)]$, near the sidereal and diurnal frequencies and harmonics. The specific frequencies that are fitted for are shown in fig. 5. Cross correlation coefficients between them vary between $10^{-5}$ to $10^{-2}$ for frequencies separated by the order of the diurnal and sidereal and are thus very well distinguished. For components centred around the sidereal frequency, and separated by frequencies of order the annual frequency, correlation coefficients vary from $10^{-3}$ to $10^{-1}$, which is still sufficiently small for independent determination at the selected frequencies.

The quadrature amplitudes with respect to the sidereal frequency are shown in fig. 5. In particular, we determine $C_{\omega_\oplus - \Omega_\oplus}$ = -1.7(1.0)×$10^{-16}$ and $S_{\omega_\oplus}$ = 0.56(1.0)×$10^{-16}$. Both amplitudes are less than two times the standard error and are not significant compared to the scatter of values calculated at nearby frequencies and harmonics. Ignoring the annual variations in (7) and taking the weighted mean of the contributions of $C_{\omega_\oplus - \Omega_\oplus}$ and $S_{\omega_\oplus}$, from (8) we obtain $P_{KT} = -1.7(4.0) \times 10^{-8}$.

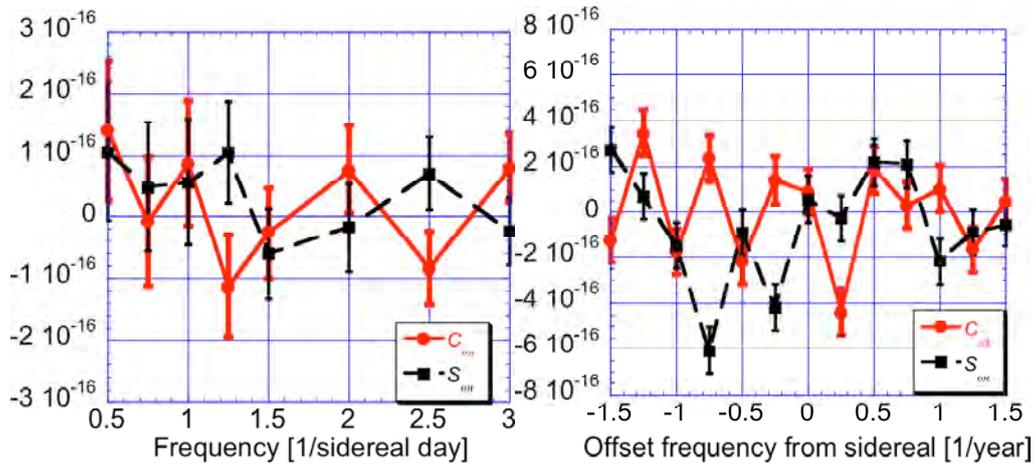

Fig. 5. Calculated coefficients $C_{\omega_t}$ and $S_{\omega_t}$ as a function of frequency. Left, over frequencies of order a sidereal day (and harmonics). Right, frequencies offset from the sidereal by the annual frequency and fractions. Error bars represent the standard error.

### B. Search for Annual Variations

To search for annual variations in the data we show here that it is optimum to search the derivative of the beat frequency data as it filters out directly the large systematic jumps between cryogenic refills and relocking of the CSO. Since, the dewar must be refilled at least once every three weeks this occurs regularly in one year. To be able to track variations of the order of a year or more it is also important that the CSO runs continuously cold without warming up to room temperature as this will change the stress conditions on the resonator as it is cycled[55]. If this is the case, the shift in the derivative of the beat frequency of the CSO will be larger than the stationary noise in the data and can mimic a positive signal when fitting harmonic amplitudes. From the data obtained since September 2002, there were only two periods, which spanned more than one year without warming up. This includes data from the 26th March 2003 to the 3rd of August 2006, and the 12th of August 2006 to the 26th of August 2008. The recorded beat frequency is shown in Fig. 6.

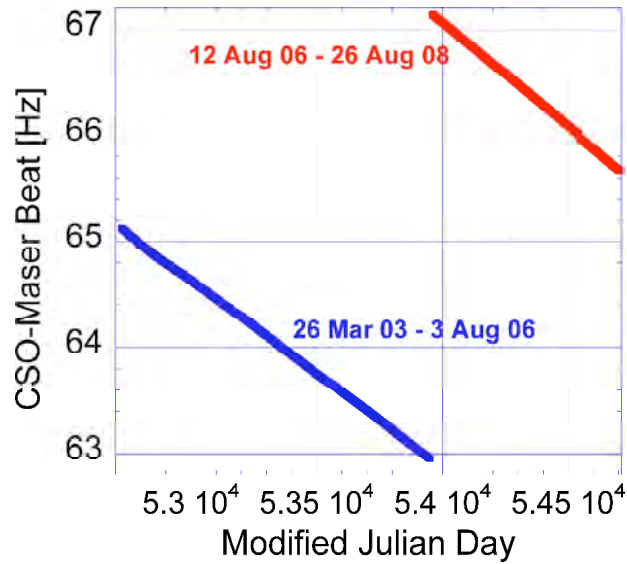

Fig. 6. The direct beat frequency between the CSO and multiplied H-maser with respect to 12 GHz.

The residuals of the beat frequency were determined by subtracting quadratic fits to the two sets of data shown in Fig. 6. During the 3$^{rd}$ of August there was a fault in the CSO, which lead to a necessary warm up from 4 K to room temperature. After fixing and re-cooling operation re-commenced on the 12$^{th}$ of August. The resulting residuals are shown in fig. 7, along with the power spectral density. Given that the power spectral density has approximately an $f^2$ dependence it is better to analyse the derivative of the beat frequency.

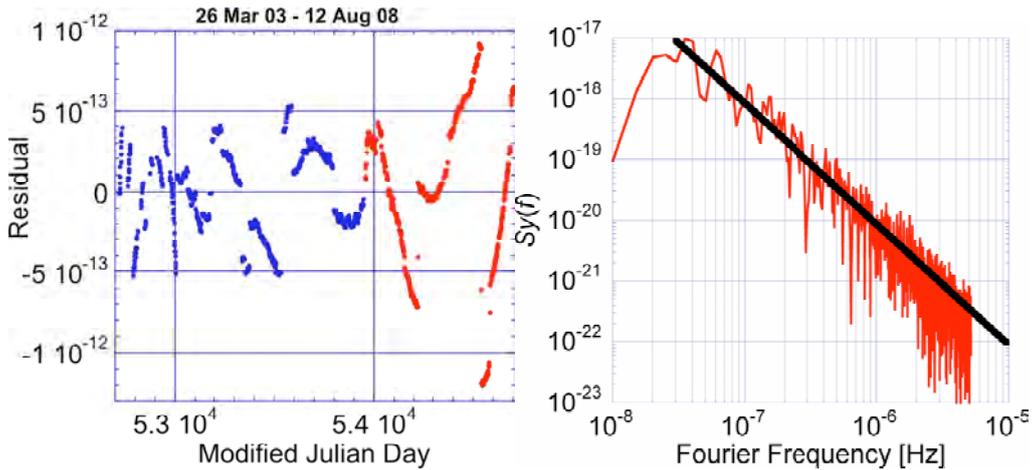

Fig. 7. Left, fractional residuals (with respect to 12 GHz) between the quadratic fit and experimental data. Right, $Sy(f)$ [Hz/√Hz], the fractional frequency power spectral density of the residuals, which shows a $f^2$ dependence.

The fractional derivative of the beat frequency (with respect to 12 GHz) was also calculated from the data and is shown in Fig. 8. The slope is approximately -1.5×10$^{-13}$/day (as we reported before)[55]. However there is clearly a small linear dependence of the derivative over the longer term, which justifies the quadratic fit to the beat frequency when calculating the residuals. Also, we note a larger than usual scatter, which relaxed after 125 days and coincided with the jump in frequency between the 3$^{rd}$ and 12$^{th}$ of August 2006 due to the necessary warm up and maintenance. It was necessary to veto this data as it is a clear systematic effect on the long-term drift rate of the resonator. Linear fits where then subtracted from both data sets (before and after the 3$^{rd}$ of August 2006), with the residuals shown in fig. 8. The power spectral density of the residuals is shown in Fig. 9, and is consistent with white noise and thus OLS was used to search for putative sinusoidal signals from the data.

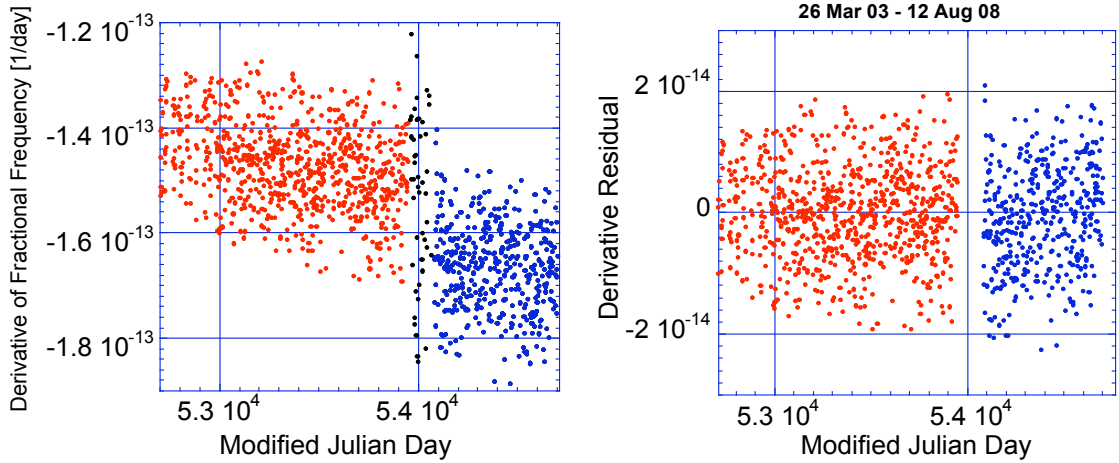

Fig.8. Left, Derivative of the fractional frequency (with outliers removed), which shows a slight variation to the linear drift of the CSO over a period of 5.5 years. Straight after maintenance (including warm up) in August 2006, there was a larger scatter in the determination of the slope, which only settled after 125 days, this was vetoed from the data and represent 6% of the total data. Right, the residuals with the linear dependence removed separately for the two data sets.

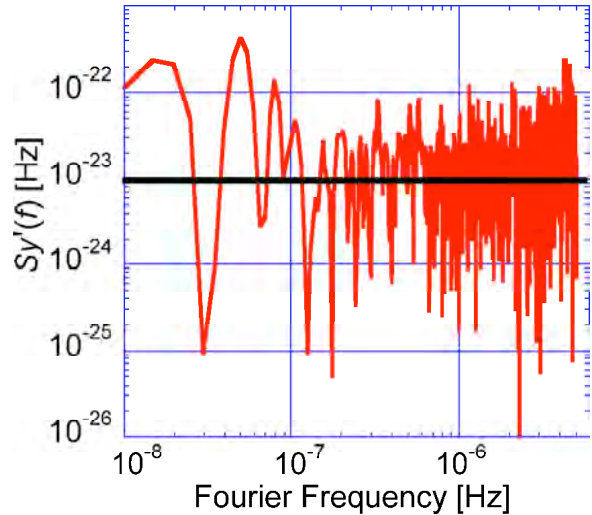

Fig. 9. Power spectral density of the residuals of the derivative of the fractional frequency fluctuations.

With respect to the beat frequency residuals shown in Fig. 7, the annual terms we search for are of the form, $C_{\Omega_\oplus} Cos[\Omega_\oplus(t-t_o)]$ (in-phase), and $S_{\Omega_\oplus} Sin[\Omega_\oplus(t-t_o)]$ (out of phase). Thus, with respect to the derivative of the beat frequency given by the residuals in Fig 8., we search for signals of the form $-\Omega_\oplus C_{\Omega_\oplus} Sin[\Omega_\oplus(t-t_o)]$ and $\Omega_\oplus S_{\Omega_\oplus} Cos[\Omega_\oplus(t-t_o)]$. In the analysis we also simultaneously search for a sum of eight nearby frequencies with the phase defined with respect to the CMB frame and of the form $\sum_i C_{\omega_i} Cos[\omega_i(t-t_o)] + S_{\omega_i} Sin[\omega_i(t-t_o)]$ from the residuals of the beat frequency (fig. 7), and of the form $\sum_i -\omega_i C_{\omega_i} Sin[\omega_i(t-t_o)] + \omega_i S_{\omega_i} Cos[\omega_i(t-t_o)]$ from the residuals of the derivative (fig. 8). Our results determine that searching the derivative of the beat frequency is about a factor of 4 more sensitive than searching the beat frequency directly. Moreover, the quadrature amplitudes using the derivative technique do not have amplitudes calculated greater than 3 standard errors over the 16 quadrature components, with all except one coefficient below two standard errors in magnitude. In contrast, when searching the residuals of the beat frequency directly, five amplitudes have significance greater than three standard errors. It is apparent that the non-stationary noise in the data can mimic significant amplitudes at many different frequencies. Thus, we should only consider searching the derivative of the beat signal when searching for putative new physics over the annual period.

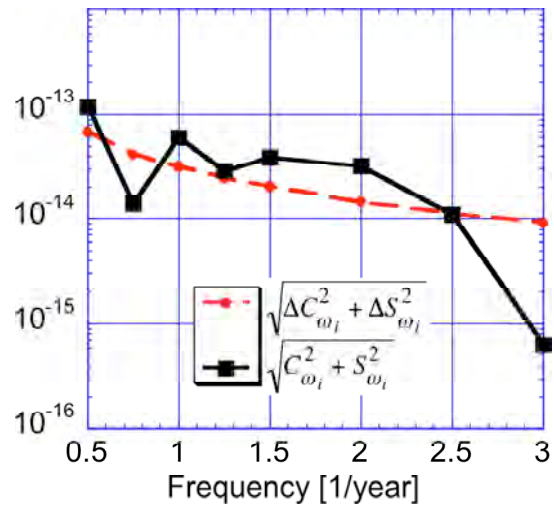

Fig. 10. OLS analysis on the residuals of the derivative shown in Fig. 8. The bold curve shows the simultaneously fitted amplitudes at various frequencies (units year$^{-1}$), while the dashed curve shows the standard error at the same frequencies.

In general, fitting the amplitudes using the derivative does not show any significance much greater than the standard error at $\Omega_\oplus$, or much greater than the amplitudes at other nearby frequencies.

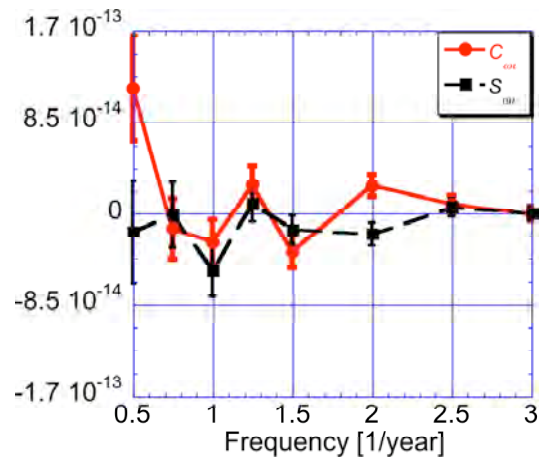

Fig. 11. Calculated coefficients $C_i$ and $S_i$ as a function of frequency over periods of order of a year.

The quadrature amplitudes at the fitted frequencies are shown in fig. 11. Cross correlation coefficients vary between $10^{-3}$ to $10^{-1}$, which is still sufficiently small for independent determination at the selected frequencies. From these results we determine $C_{\Omega_\oplus}$ = -2.7(2.1)×$10^{-14}$ and $S_{\Omega_\oplus}$ = -5.4(2.4)×$10^{-14}$. Combining (7) and (8) with this result we can put a limit on $P_{KT}$ of -2.3(1.0)×$10^{-7}$.

### III. LOCAL POSITION INVARIANCE

### B. Search for Diurnal Variations

With respect to the frequency comparison between the CSO and H-maser given by (10), we also search for signals in phase, $C_{\omega_\oplus - \Omega_\oplus} Cos[(\omega_\oplus - \Omega_\oplus) t_{noon}]$, and out of phase, $S_{\omega_\oplus - \Omega_\oplus} Sin[(\omega_\oplus - \Omega_\oplus) t_{noon}]$, with the varying gravitational potential. Here $\omega_\oplus - \Omega_\oplus$ is the diurnal frequency ($2\pi$ rads/day), $C_{\omega_\oplus - \Omega_\oplus}$ the in phase and $S_{\omega_\oplus - \Omega_\oplus}$ the out of phase quadrature amplitudes. Zero phase (or $t_{noon} = 0$) is set when Paris is closest to the sun of the first day of all the data sets (9$^{th}$ of September). Thus, with respect to the derivative of the beat frequency the quadrature amplitudes are determined by searching the residuals for signals of the form; $-(\omega_\oplus - \Omega_\oplus) C_{\omega_\oplus - \Omega_\oplus} Sin[(\omega_\oplus - \Omega_\oplus) t_{noon}]$ and $(\omega_\oplus - \Omega_\oplus) S_{\omega_\oplus - \Omega_\oplus} Cos[(\omega_\oplus - \Omega_\oplus) t_{noon}]$. We use the same process as the Kennedy-Thorndike experiment and simultaneously fit 40 parameters simultaneously near the diurnal frequency and at harmonics. The calculation of the $C_{\omega_i}$ and $S_{\omega_i}$ coefficients are very similar to those in Fig. 5, but just in a different phase and return a value of $C_{\omega_\oplus - \Omega_\oplus}$ = 1.8(1.0)×$10^{-16}$ and $S_{\omega_\oplus - \Omega_\oplus}$ = 2.0(1.0)×$10^{-16}$. Thus, combining these values with (10) and dividing the in-phase coefficient by the variation of the gravitational potential due to the Earth's rotation (-2.54×$10^{-13}$), a limit on LPI and fundamental constants can be determined to be; $\beta_{H\text{-}MASER}$-$\beta_{CSO}$ = $3\kappa_\alpha + \kappa_e - 0.1\kappa_q$ = -6.9(4.0)×$10^{-4}$. Note that only the in-phase term matters when looking for dependence proportional to ambient gravitational field (eq. (10)). The quadrature is just a "check" similar to looking at neighbouring frequencies.

### A. Search for Annual Variations

We search for signals in phase with the varying gravitational potential, and of the form $-\Omega_\oplus C_{\Omega_\oplus} Sin[\Omega_\oplus t_\oplus]$ from the data presented in fig. 8. We also search for the quadrature amplitude of the form $\Omega_\oplus S_{\Omega_\oplus} Cos[\Omega_\oplus t_\oplus]$ as well as the in-phase and quadrature amplitudes at seven other frequencies shown in fig. 10 and 11. The determined coefficients are of similar values but differ due to the phase being set by the time of an Aphelion rather than the J2000 equinox used for the boost tests previously. In general, the amplitudes at $\Omega_\oplus$ do not show any significance much greater than the standard error, and much greater than the other nearby frequencies and return a value of $C_{\Omega_\oplus}$ = 4.5(2.4)×10$^{-14}$ and $S_{\Omega_\oplus}$ = -3.9(2.1)×10$^{-14}$. Dividing the in-phase coefficient by the amplitude of the varying gravitational potential given by (11) (-1.65×10$^{-10}$) we derive the following limit with respect to the varying gravitational potential; $\beta_{H-maser} - \beta_{CSO}$ = $3\kappa_\alpha + \kappa_e - 0.1\kappa_q$ = $-2.7(1.4) \times 10^{-4}$. The constraint on the electromagnetic resonator ($\beta_{H-Maser} - \beta_{CSO}$) is two orders of magnitude better than the previous result of Turneaure et. al.[11] of 1.7x10$^{-2}$. Such tests on electromagnetic resonators would be sensitive to Lorentz violations that couple the gravity sector to the photon. Otherwise, the constraint with respect to fundamental constants is one to two orders of magnitude worse as those constrained recently by atomic clocks[14, 16, 17].

## IV BOOST DEPENDENCE OF FUNDAMENTAL CONSTANTS

To obtain a limit on the components of the vector **B** as defined in (15) the amplitude due to the violation is calculated by taking the dot product of the boost given in (5) with **B**, and returns components at the annual and sidereal frequencies as shown in Table I. There are four measurements made with three amplitudes to constrain ($B_x$, $B_y$, $B_z$), thus the system is over constrained and limits on all three components may be made.

Table I. The amplitude dependence on coefficients ($B_x$, $B_y$, $B_z$), due to the putative boost dependence of the fundamental constants (occurring at annual and sidereal frequencies). To the right of column 2 the numeric values and standard errors are given for each component, which have been determined experimentally.

| Frequency Component | Amplitude | |
|---|---|---|
| $\sin[\Omega_\oplus(t - t_o)]$ | $9.71\times10^{-5} B_x + 2.08\times10^{-5} B_y$ | -5.4 (2.4) ×10$^{-14}$ |
| $\cos[\Omega_\oplus(t - t_o)]$ | $1.91\times10^{-5} B_x - 8.92\times10^{-5} B_y + 3.92\times10^{-5} B_z$ | -2.7 (2.1) ×10$^{-14}$ |
| $\sin[\omega_\oplus(t - t_o) + \Phi]$ | $-1.02\times10^{-6} B_x$ | 0.56 (1.0) ×10$^{-16}$ |
| $\cos[\omega_\oplus(t - t_o) + \Phi]$ | $1.02\times10^{-6} B_y$ | 0.87 (1.0) ×10$^{-16}$ |

The system of equations may be solved using weighted averaging for an over determined equation set, with the weightings set by the inverse variance (square of the standard errors in table I). The process returns values of ($B_x$, $B_y$, $B_z$) of (-1.3(0.9)×10$^{-10}$, 0.6(1.0)×10$^{-10}$, -4.7(4.8)×10$^{-10}$), allows the first upper limit of variation of fundamental constants to boost with respect to the CMB.

## V DISCUSSION

In our experiment, the LPI test with respect to gravitational potential variation over the annual period is a factor of 3 more sensitive than the diurnal. In contrast, the LLI and fundamental constant test with respect to boost effects over the sidereal and diurnal periods is a factor of 2.5 times more sensitive than the annual. Thus, combining the results as a weighted mean may attain a small improvement in the determination of the values. Doing this sets following limits; 1) $\beta_{H-maser} - \beta_{CSO} = 3\kappa_\alpha + \kappa_e - 0.1\kappa_q = -3.2(1.3)\times10^{-4}$, 2) $P_{KT} = -4.8(3.7)\times10^{-8}$ and 3) $\mathbf{B} = 3\mathbf{B}_\alpha + \mathbf{B}_e - 0.1\mathbf{B}_q = (-1.3(0.9)\times10^{-10}, 0.6(1.0)\times10^{-10}, -4.7(4.8)\times10^{-10})$. The first result improves LPI gravitational tests on electromagnetic resonators by two orders of magnitude, the second result improves Kennedy-Thorndike experiments by a factor of eight, while the final result puts the first determination on the variation of fundamental constants with respect to boost.

It is difficult to experiment on systematics over a year period because of the large time scales involved. For example, limits using atomic clock comparisons typically estimate constant systematic shifts from the absolute frequency over much smaller time scales, following this averaged results are presented with an error budget over a many periods smaller than one year. Our approach is different as we look at the relative stability between the two oscillators and do a direct investigation using least squares to search for the amplitudes at the required frequencies. For this type of experiment a constant systematic shift (or shifts at frequencies other than the frequencies of interest) does not limit the measurement. One just needs to show that a systematic is not limiting the measurement at the required frequencies of interest. Typical possible systematics include; frequency dependence on pressure, temperature, tilt, magnetic field etc[19]. For our purposes these effects must couple to variations over the annual, sidereal and diurnal time scales. It is feasible that a systematic with a certain phase may be present due to day-night and seasonal variations in these parameters. If this was indeed the case one would expect to see a significant signal much greater than the standard error and much larger than nearby frequencies. For this experiment it is not the case at our frequencies of interest. Given that the diurnal frequency is equal to the difference between the sidereal and annual frequency, to provide a more thorough check, we have searched for 40 coefficients simultaneously for amplitudes offset from the sidereal by the annual frequency as shown in fig. 5. There was no significant difference at the annual offsets from the sidereal frequency. Also, we searched for amplitudes at frequencies of fractions and harmonics of the annual frequency as shown in fig. 10. The combined result indicates that there is likely only a small systematic influence at the diurnal, sidereal and annual frequencies since we record null results.